\documentclass[twocolumn,pra,showpacs]{revtex4}

\usepackage{amsmath}
\usepackage{graphics}
\usepackage{epsfig}

\begin{document}

\title{Manifestations of the Efimov Effect for Three Identical
  Bosons} 
\author{J. P. D'Incao and B. D. Esry}
\affiliation{Department of Physics, Kansas State University,
Manhattan, Kansas 66506} 

\begin{abstract}
In this paper we present results from numerical calculations for
three identical boson systems for both very large and infinite  
two-body $s$-wave scattering length $a$. We have considered scattering
lengths up to $2\times 10^5$~a.u. and solved the hyperangular part of the
Schr\"odinger equation for distances up to $10^6$~a.u.. Form these, we
obtained the three-body effective potentials, hyperspherical channel
functions and the asymptotic behavior of the nonadiabatic couplings in
order to to characterize the main aspects of the Efimov states. These
results allow us to test and quantify the assumptions related to the Efimov
effect.
\end{abstract}
\maketitle

\section{Introduction}

The experimental ability to create ultracold atomic systems, such as
Bose-Einstein condensates, with a large range of interaction strengths
has raised the possibility of investigating a very unusual phenomenon:
the Efimov effect. The Efimov effect \cite{Efimov} is responsible
for the emergence of an attractive long-range effective three-body
interaction irrespective of the short-range character of the atomic
interactions, producing a large 
number of three-body bound states --- called Efimov states --- even
when the two-body subsystems have none. This effect is expected to occur
in the limit where the two-body $s$-wave scattering length $a$ is much
larger than the characteristic range $r_{0}$ of the interatomic 
interactions. Such a regime can be reached, for instance, by 
tuning the two-body interactions via an external magnetic 
field near a Feshbach resonance
\cite{Inouye,Courteille,Roberts,Kevrekidis,Weber}. Besides its
intriguing impact on the three-body bound state spectrum, Efimov
physics has been invoked to explain the remarkable universality of
three-body scattering observables at ultracold energies
\cite{Esry-01,NielsenEsry,Braaten,BraatenReview,Macek-02,Dincao-I,ScaLenLaws,Bulgac}.
Here, we use the less restrictive term ``Efimov physics'' to indicate
the qualitative change in behavior exhibited by the system when
$|a|\rightarrow\infty$.  

Despite their importance, neither Efimov states nor any other
signature of Efimov physics has yet been detected. The main difficulty
for direct observation of these states stems from the fact that the
Efimov states are very weakly bound and easily perturbed. 
Indirect observation of Efimov physics might be possible, however, by
measuring some three-body scattering observables like three-body
recombination or vibrational relaxation
\cite{Esry-01,Macek-02,NielsenEsry,Braaten,BraatenReview,Dincao-I} as
a function of the scattering length. The detection of the series
of interference minima and resonance peaks present in the three-body
recombination rate as a function $a$, or of the resonance peaks
in the vibrational relaxation rate, would be a clear signature of 
the Efimov physics involved in the ultracold three-body
collisions. In this case, the main difficulties arise from the fact
that finite collision energies --- even ultracold energies --- reduce
the range of $a$ in which Efimov physics can be observed
\cite{Dincao-I}. At finite energies, the unitary limit, high partial
waves, and thermal averaging tend to wash out the features connected
to Efimov physics. The coldest current experiments, however, might
offer a large enough window in scattering length to see Efimov
physics, but they would still be difficult experiments. For
Bose-Eistein condensates of $^{4}$He, $^{23}$Na and $^{87}$Rb, for
instance, we can expect to observe at most three interference minima
or resonance peaks in the three-body recombination rate for energies
of $3$~nK, $0.5$~nK and $0.1$~nK, respectively \cite{Dincao-I}. To see
more than three of either requires even lower temperatures.  

Previous theoretical work
\cite{Esry-01,Macek-02,NielsenEsry,Braaten,BraatenReview,Dincao-I}
has revealed many universal low-energy three-body scattering
properties related to Efimov physics. In this regime, the effective
interactions and the three-body observables are determined primarily
by $a$, independent of the details of the two-body interactions. 
The assumption that the attractive long-range behavior of the
three-body effective potential ($\propto -1/R^{2}$) is valid in the
range $r_{0}\ll R\ll |a|$, with $R$ giving the overall size of the
three-body system, is usually required in order to determine the $a$
dependence in the three-body obsevables. 
In fact, the more relaxed assumption that this 
potential holds even for $r_{0}\lesssim R \lesssim |a|$ has been used
to deduce the number of Efimov states and other low-energy three-body
observables \cite{Macek-02,Braaten,BraatenReview,Bulgac}.
Neither assumption, however, has been thoroughly tested against
accurate calculations. Given the central role these assumptions play
in determining the universal properties of ultracold three-body
scattering, they should be tested.

For this paper, we have solved the Schr\"odinger equation for three
identical bosons using the adiabatic hyperspherical representation for
the $J^{\pi}=0^{+}$ symmetry over a broad range of scattering lengths,
including essentially infinite values. The numerical results were
obtained in a regime not accessible in previous calculations, allowing
us to test the assumptions above by checking the modifications to the
effective three-body potentials as $|a| \rightarrow \infty$. In
addition, we have explored the validity of the above assumptions as
the number of two-body bound states in the two-body potential is
increased. We have considered both positive and negative scattering
lengths, up to $2\times 10^5$~a.u. in magnitude and separately the
infinite $a$ case. 
We solved the three-body Schr\"odinger equation for
distances $R$ as large as $10^6$~a.u. in order to determine the
effective three-body potentials as well as the nonadiabatic couplings
responsible for inelastic transitions. The asymptotic behavior of the
coupling and the hyperspherical channel functions was investigated to
point out the main characteristics of the Efimov states in the
adiabatic hyperspherical point of view.

\section{Adiabatic Hyperspherical Representation}

We solve the three-body Schr\"odinger equation using the
adiabatic hyperspherical representation \cite{Esry-02,Esry-03}. After 
the usual separation of the center-of-mass motion, the
three-body problem can be described by the hyperradius $R$ and five 
angles, denoted collectively by $\Omega$. 
The five angular coordinates are chosen to be
the Euler angles ($\alpha$, $\beta$ and $\gamma$) that specify the
orientation of the the plane defined by the three particles relative
to the space-fixed frame, plus two hyperangles ($\varphi$ and
$\theta$) defined as a modification of Smith-Whitten
coordinates \cite{Esry-02,Esry-03,Smith}. 

The Schr\"odinger equation in hyperspherical coordinates can be
written in terms of the rescaled wave function 
$\psi=R^{5/2}\Psi$ (in atomic units),
\begin{eqnarray} 
\left[-\frac{1}{2\mu}\frac{\partial^2}{\partial R^2}
+H_{\rm ad}(R,\Omega)\right]\psi(R,\Omega)=E\psi(R,\Omega),\label{schr} 
\end{eqnarray} 
\noindent
where $\mu$ is the three-body reduced mass (we choose $\mu=m/\sqrt{3}$
for three identical bosons of mass $m$) and $E$ is the total
three-body energy. The adiabatic Hamiltonian $H_{\rm ad}$ is given by
\begin{eqnarray}
H_{\rm ad}(R,\Omega)=\frac{\Lambda^{2}(\Omega)+\frac{15}{4}}{2\mu
  R^2}+V(R,\varphi,\theta), 
\label{had}
\end{eqnarray} 
\noindent
in which $\Lambda^{2}$ is the grand angular momentum
operator and $V$ contains the interparticle interactions.

In the adiabatic hyperspherical representation, the total wave
function is expanded on the channel functions $\Phi_{\nu}(R;\Omega)$,
\begin{equation}
\psi(R,\Omega)=\sum_{\nu}F_{\nu}(R)\Phi_{\nu}(R;\Omega),\label{chfun}
\end{equation}
\noindent
where the expansion coefficient $F_{\nu}(R)$ is the hyperradial wave
function. The total wave function is still, in principle, exact. The
channel functions $\Phi_{\nu}(R;\Omega)$ form a complete set of
orthonormal functions at each $R$ and are eigenfunctions of the
adiabatic Hamiltonian: 
\begin{equation}
H_{\rm ad}(R,\Omega)\Phi_{\nu}(R;\Omega)
=U_{\nu}(R)\Phi_{\nu}(R;\Omega).\label{poteq}
\end{equation}
\noindent
The eigenvalues of this equation, $U_{\nu}(R)$, play the role of
effective potentials for the hyperradial motion. This fact can be seen
upon substitution of Eq.~(\ref{chfun}) into the Schr\"odinger
equation (\ref{schr}) and projecting out $\Phi_{\nu}$. The result is a
system of coupled ordinary differential equations 
\begin{widetext}
\begin{eqnarray}
\left[-\frac{1}{2\mu}\frac{d^2}{dR^2}+U_{\nu}(R)\right]F_{\nu}(R) 
-\frac{1}{2\mu}\sum_{\nu'}
\left[2P_{\nu\nu'}(R)\frac{d}{dR}+Q_{\nu\nu'}(R)\right]F_{\nu'}(R)
=EF_{\nu}(R),\label{radeq}
\end{eqnarray}
\end{widetext}
\noindent
where $P_{\nu\nu'}(R)$ and $Q_{\nu\nu'}(R)$ are the nonadiabatic
coupling terms. These couplings are generated by the $R$-dependence of
the channel functions and are responsible for the inelastic 
transitions in three-body scattering. They are defined as 
\begin{eqnarray} 
P_{\nu\nu'}(R) &=&
\Big\langle\hspace{-0.15cm}\Big\langle\Phi_{\nu}(R)\Big|
\frac{d}{dR}\Big|\Phi_{\nu'}(R)\Big\rangle\hspace{-0.15cm}\Big\rangle
\label{puv}
\end{eqnarray}
\noindent
and
\begin{eqnarray} 
Q_{\nu\nu'}(R) &=&
\Big\langle\hspace{-0.15cm}\Big\langle\Phi_{\nu}(R)\Big|
\frac{d^2}{dR^2}\Big|\Phi_{\nu'}(R)\Big\rangle\hspace{-0.15cm}\Big\rangle,
\label{quv}
\end{eqnarray} 
\noindent
where the double brackets denote integration over the angular coordinates
$\Omega$ only. Here, we will concentrate the solutions of the
adiabatic equation (\ref{poteq}), namely the potential curves and
couplings. 
For the present calculations, the potential $V$ in
Eq.~(\ref{had}) is a pairwise sum of short-range potentials,
$V=v(r_{12})+v(r_{31})+v(r_{23})$, which is a good model for
spin-stretched atoms.

\subsection{Two-body potential model}

To a good approximation, the Efimov effect, as well as other low-energy
three-body properties, do not depend on the details of the interatomic
interaction, only on the scattering length and $r_{0}$, the
characteristic range of the interatomic interactions. This
approximation, verified for different two-body potentials
\cite{Esry-01}, allows the use of model potentials, substantially
simplifying the numerical calculations by controlling the number of
two-body channels in the problem. 
The two-body potential adopted here is
$v(r)=D\mbox{sech}^2(r/{r_{0}})$. With this potential, we can easily
produce two-body systems with different values of the scattering
length by changing the interaction strength $D$. The scattering length
is determined for each value of $D$ by solving the two-body scattering
problem. For infinite values of the scattering length, however, we
used a slightly different approach which will be discussed in more
detail below. We have considered potentials
with just one two-body bound state and a characteristic range of
$r_{0} = 15$~a.u., and using the mass of $^{4}$He atoms.  

\begin{figure}[htbp]
\includegraphics[width=2.2in,angle=270]{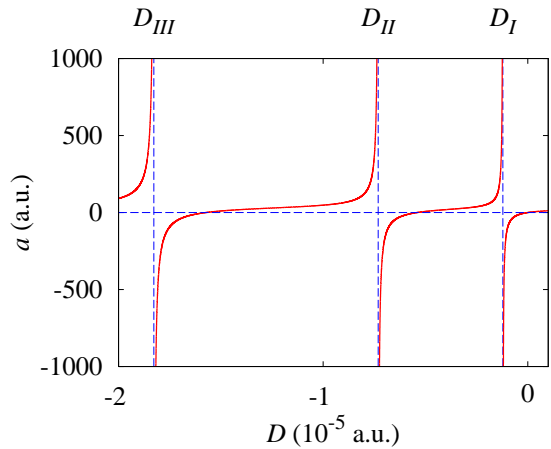}
\caption{The two-body $s$-wave scattering length $a$ as a function of
  the potential strength $D$. When the two-body potential is deep
  enough to support an additional bound state, $a$ passes through a
  pole, indicated by $D_{I}$, $D_{II}$ and $D_{III}$. \label{fig.a}}  
\end{figure}

Figure \ref{fig.a} shows the scattering length $a$ as a
function of the potential strength $D$. 
Each time that the two-body potential deepens enough
to support an additional $s$-wave bound state, the scattering length
goes through a pole, labeled in Fig.~\ref{fig.a} by  $D_{I}$, $D_{II}$
and $D_{III}$. For $D>D_{I}$, the two-body system has no bound state
and represents the standard Efimov case when $|a| \rightarrow
\infty$. For $D_{II}<D<D_{I}$, the system has a single $s$-wave bound
state, and for $D_{III}<D<D_{II}$,
the system has gained another $s$-wave bound state. 
Our model two-body potential mimics the magnetic field control
of the scattering length near a two-body $s$-wave Feshbach resonance
\cite{Inouye,Courteille,Roberts,Kevrekidis,Weber}, assuming that the
only effect the resonance is to change the scattering
length. In fact, we expect this model to be valid in the threshold
regime, i.e., when the collision energy is the smallest energy in the
system \cite{Dincao-I}.
As $a \rightarrow +\infty$, the two-body bound state is weakly
bound, and the three-body effective potential that approaches this
state asymptotically supports the three-body Efimov states. As
$a \rightarrow -\infty$, the two-body bound state becomes a deep
bound state and the Efimov states are resonances rather than bound
states.  
\noindent

To determine the value of $D$ that gives a particular scattering
length, we simply read the value from Fig. \ref{fig.a}. To determine
the value of $D$ that gives $|a| \rightarrow \infty$ from the figure,
though, is not 
simple to do accurately. We thus used a slightly different approach
that takes advantage of the fact that the derivative of the
two-body radial wave function must vanish at large distances
when $|a| = \infty$. To this end, we solved the two-body radial
Schr\"odinger equation at zero energy,
\noindent
\begin{equation}
\left[-\frac{1}{m}\frac{d^2}{dr^2}+D\mbox{sech}^2(r/r_{0})\right]u(r)=0,
\label{2b-eq} 
\end{equation}
\noindent
and required $u'(r)=0$ at some large distance. The potential strength
$D$ is the eigenvalue of Eq.~(\ref{2b-eq}) --- the lowest corresponds
to the solution with no nodes $D=D_{I}$, the next to the solution with
one node $D=D_{II}$, and so on. This procedure determines $D$ to
essentially machine precision.  
We can then perform three-body calculations with a scattering length
as close as numerically possible to infinity. 

It is worth to noting that the solutions of the $\mbox{sech}^2$ potential are
known analytically, so that we could, in principle, determine these
values of $D$ exactly. We used the present method because it will work
for any short-range potential.

\subsection{Effective hyperspherical potentials}

Neglecting all off-diagonal terms in Eq.~(\ref{radeq}) defines the
effective hyperspherical potential, 
\begin{eqnarray} 
W_{\nu}(R)=U_{\nu}(R)-\frac{1}{2\mu}Q_{\nu\nu}(R).\label{effpot}
\end{eqnarray} 
\noindent
For short-range two-body potentials the channel index $\nu$ labels
both two-body and three-body configurations in the asymptotic limit
($R \rightarrow \infty$). In this limit, the potential curves
$U_{\nu}(R)$ approach either a two-body bound state energy for the  
bound channels (BC) or zero energy for the three-body continuum
channels (CC). In both cases, $Q_{\nu\nu}$ vanishes for
$R \rightarrow \infty$. 

For finite $a$, however, the effective potentials are influenced by
Efimov physics in the range $r_{0}$$\ll$$ R$$ \ll $$|a|$. For these
$R$, the effective potentials areproportional to $R^{-2}$, and can be
attractive or repulsive. The channel of primiry interest in this case
is the one that leads to the Efimov effect --- for $a>0$, it is the
chanel correlated asymptotically with the weakly bound $s$-wave dimer;
for $a<0$, it is the lowest continuum channel. Because of its
importance in our discussion we label it the ``Efimov channel'' (EC). 
It is worth to mention that for $R\gg|a|$ the Efimov channel behaves
as a bound or continuum channel, as appropriate.

\begin{figure}[htbp]
\includegraphics[width=3.8in,angle=270]{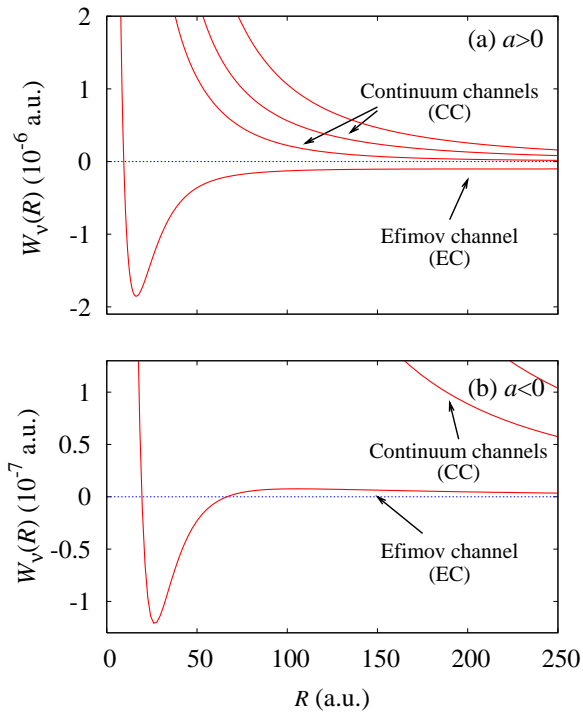}
\caption{The three-body effective potential for (a) $a>0$ and (b)
  $a<0$, calculated near pole $D=D_{I}$. For $a>0$, the system has
  only one bound state, while for $a<0$ the system has none.
\label{fig.b.new}}  
\end{figure}

In Fig.~\ref{fig.b.new} we show the effective potentials
$W_{\nu}(R)$ calculated near the pole $D=D_{I}$
in Fig.~\ref{fig.a}. For $a>0$, the lowest potential is the Efimov 
channel (EC), converging to the two-body bound state for
$R\rightarrow\infty$. The higher lying potentials are associated with
three-body continuum channels (CC). 
As $D$ is increased towards $D_{I}$ from the left,
$a\rightarrow+\infty$ and the dimer binding energy vanishes. At the
same time, the Efimov channel evolves from that shown in
Fig.~\ref{fig.b.new} to a purely attractive $R^{-2}$ for $R\gg r_{0}. $
Similarly, for $a<0$ in Fig.~\ref{fig.b.new}(b), the lowest
potential is the Efimov channel (EC) converging to the three-body 
continuum limit as $R\rightarrow\infty$. 
In this case, $a\rightarrow-\infty$ as $D$ approaches $D_{I}$ from the
right, and the potential barrier in the Efimov channel moves to
infinity ($R\propto |a|$). At $D=D_{I}$, the potential is the same
attractive $R^{-2}$ obtained by approching $D_{I}$ from the left. Note
that for both $a<0$ and $a>0$, there can be bound channels (BC) that
lie at lower energies.
\noindent

The asymptotic behavior, $R\gg |a|$, of the effective potential
$W_{\nu}(R)$ can be determined analytically \cite{Macek-03,Nielsen}.  
For the bound channels (BC), the asymptotic limit represents the
atom-dimer configuration, and the asymptotic effective potential is
given by 
\begin{eqnarray} 
W_{\nu}(R)=E_{vl'}+\frac{l(l+1)}{2\mu R^2}+
{\cal O}\left(\frac{1}{R^3}\right).\label{boundch}
\end{eqnarray} 
\noindent
The rovibrational energy $E_{vl'}$ of the dimer is labeled by its
vibrational quantum number $v$ and its angular momentum $l'$; 
$l$ is the relative angular momentum between the dimer and
the third atom. The notation ${\cal O}(...)$ indicates the leading
order correction. 

The three-body continuum channels (CC) represent three free atoms,
each one far from the others in the asymptotic limit. The two-body
potentials thus vanish, leaving only the kinetic energy. Therefore,
the leading term of the effective potential is determined by the
eigenvalues $\lambda(\lambda+4)$ of the kinetic energy, and the higher order
terms can be written in terms of the scattering length $a$, 
\begin{eqnarray} 
W_{\nu}(R)=\frac{\lambda(\lambda+4)+\frac{15}{4}}{2\mu R^2}
+\frac{c_{1}}{\mu R^2}\left(\frac{a}{R}\right)^{2l_{0}+1}\nonumber \\
+\frac{c_{2}}{\mu R^2}\left(\frac{a}{R}\right)^{2l_{0}+2}+
{\cal O}\left(\frac{1}{R^{2l_{0}+5}}\right),\label{contch}
\end{eqnarray}
where $c_{1}$ and $c_{2}$ are positive constants. 
We can relate the quantum number $\lambda$ to the partial angular
momenta of the system using $\lambda=2n+l+l'$ --- $n$ is a
non-negative integer, $l'$ is the angular momentum associated with the
Jacobi vector connecting particles $1$ and $2$ and $l$ with the Jacobi
vector connecting particle $3$ with the center of mass of $1$ and $2$
[see also Eq.~(\ref{boundch})]. For identical bosons with
$J^{\pi}=0^+$, symmetry requires $\lambda=0$, $4$, $6$, ..., and
$l=l'=0$, $2$, $4$, .... The quantum number $l_0$ in
Eq.~(\ref{contch}) can thus be thought of as the dominant partial wave
in the expansion of a continuum state labeled by $\lambda$. It is
non-zero only when there are degeneracies in $\lambda$ --- which begin
in the present case with $\lambda=12$. Thus, this situation does not
occur for the lowest continuum channel which is usually of most
interest for ultracold collisions. We will use $\lambda$ and $l_{0}$  
to label the three-body continuum channels
 
As discussed above for $|a|\rightarrow\infty$, the effective
potentials for both $a<0$ and $a>0$ in the range $r_{0}$$\ll$$ R$$ \ll
$$|a|$ are proportional to $R^{-2}$. 
Explicitly, 
\begin{eqnarray} 
W_{\nu}(R)=-\frac{s_{0}^2+\frac{1}{4}}{2\mu R^2}+{\cal
  O}\left(\frac{1}{R^3}\right),\label{longrange} 
\end{eqnarray} 
\noindent
with $s_{0}=1.00624$, therefore producing an infinite number of
three-body bound states when $|a|=\infty$ --- which is precisely the 
Efimov effect. The higher-lying continuum channels also feel the
influence of the Efimov physics and are no longer associated simply
with the eigenvalues of the kinetic energy. For $r_{0}$$\ll$$ R$$ \ll
$$|a|$, they are given by   
\begin{eqnarray} 
W_{\nu}(R)=\frac{s_{\nu}^2-\frac{1}{4}}{2\mu R^2}+{\cal
  O}\left(\frac{1}{R^3}\right),\label{longrangerep} 
\end{eqnarray}
\noindent
where $s_{\nu}$ is a constant that represents a sort of correlation
in addition to the kinetic energy. 

In some way, the assumption that the potentials (\ref{longrange}) and
(\ref{longrangerep}) hold for $r_{0}\ll R\ll |a|$, has been used
\cite{Macek-02,Braaten,BraatenReview,Bulgac} to deduce many properties
of three-body systems with a large scattering length.
In fact, it was the more relaxed assumption that the long-range
effective potential (\ref{longrange}) holds even for $r_{0} \lesssim R
\lesssim |a|$ that was used in
Refs.~\cite{Macek-02,Braaten,BraatenReview,Bulgac} 
to deduce some low-energy three-body observables and the number of
Efimov states, 
\begin{equation}
N\simeq(s_{0}/\pi)\ln(|a|/r_{0}).\label{NumEfimovStates}
\end{equation} 
\noindent
The validity of the assumption that  $r_{0} \lesssim R \lesssim |a|$
sets limits on the quantitative aspects of the these predictions. In
the next section, the above assumptions will be tested with accurate
numerical calculations for scattering lengths up to $200000$~a.u. (and
$r_{0} = 15$~a.u.), which is unambiguously in the regime $|a| \gg
r_{0}$.   

\section{Results and discussion \label{secIII}}

The adiabatic equation (\ref{poteq}) has been solved for hyperradii up
to $2\times 10^6$~a.u. in order to reach 
the asymptotic regime for the effective potential and couplings. We
can thus compare our numerical results with the common assumptions
of the Efimov effect. The main thrust of this investigation is
to establish quantitative bounds on the range where the Efimov
potential accurately approximates the effective potential. In
addition, we have investigated the qualitative differences between 
Efimov states and ordinary three-body bound states by analyzing the
channel function and nonadiabatic couplings when
$|a| \rightarrow \infty$. 

\subsection{Effective Potentials}

\begin{figure}[htbp]
\includegraphics[width=3.9in,angle=270]{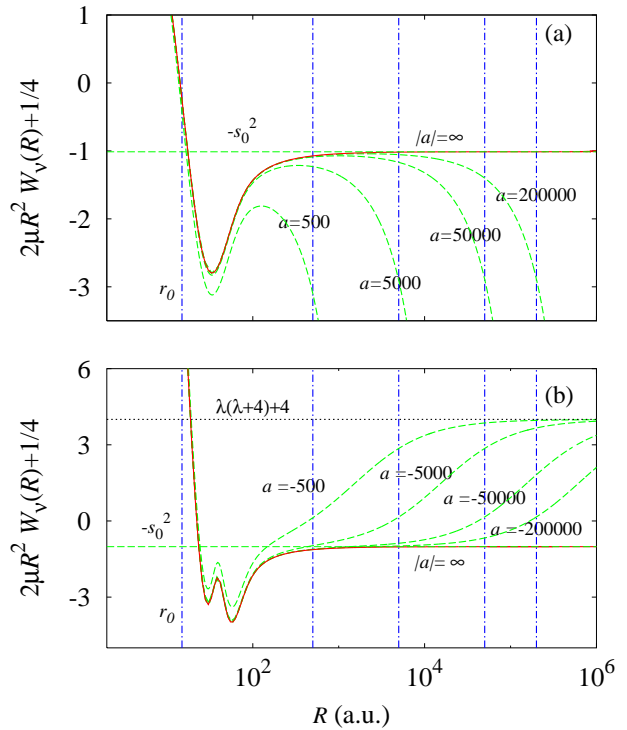} 
\caption{Three-body effective potentials for (a) $a>0$ and (b) $a<0$,
  with $|a|=500$, $10000$, $100000$, and $200000$~a.u. (dashed
  lines). Also shown is the numerical result for $|a|=\infty$ (solid
  line). The horizontal dashed line at $-s_{0}^2$ is the Efimov
  potential, Eq.~(\ref{longrange}), and the vertical lines correspond
  to $R=r_{0}$ and $R=|a|$. \label{fig.b}}    
\end{figure}

We show in Fig.~\ref{fig.b} the effective potentials associated with
the weakly bound dimer for $a>0$, and with lowest continuum channel
for $a<0$, for several values of $a$.
They are multiplied by the factor $2\mu R^2$ in order to
more clearly reveal whether they match Eq.~(\ref{longrange}). If
Eq.~(\ref{longrange}) holds, then these curves should approach the
constant $-s_{0}^{2}$($=1.0125$). Figure~\ref{fig.b} shows our
numerical potentials for $|a|=\infty$ (solid lines) and finite $a$
(dashed lines) as well. For $a>0$ 
in Fig~\ref{fig.b}(a), the effective potentials correspond to the
weakly bound channel, converging to the energy of the weakly bound
dimer state [Eq.~(\ref{boundch})] given approximately by
$-1/ma^2$. This fact manifests itself at large $R$ in Fig.~\ref{fig.b}  
by the $-R^2$ divergence of the curves. As the
scattering length grows in magnitude, the effective potential should
converge to the Efimov potential (indicated by the horizontal
dashed line) in the range $r_{0} \ll R \ll |a|$.

Similarly, Fig. \ref{fig.b}(b) shows the $a<0$ 
effective potentials corresponding to the lowest continuum channel and
converging to the Efimov potential when $|a| \rightarrow \infty$
(horizontal dashed line). For finite $a$, this curve must
asymptotically approach the lowest kinetic energy eigenvalue
(dotted line) --- $\lambda(\lambda+4)+4$ with
$\lambda=0$. The lowest effective potential converges to the deeply 
bound dimer state and is omitted. Efimov states in this case are thus 
resonances rather than bound states, since they can decay to the
lower-energy dimer channels 
\cite{NielsenEsry}. 

The vertical lines in Fig.~\ref{fig.b} correspond to $R=r_{0}$
and $R=|a|$, with $|a|=500$, $10000$, $100000$ and
$200000$~a.u.. Within this region --- actually $r_{0}\ll R\ll |a|$ 
--- the associated potential is expected to be universal, i.e., behave
like the Efimov potential. Clearly, the potentials do not satisfy this
expectation over the whole range of $r_{0} \le R \le |a|$ as is often
assumed. Instead, the condition specifies $R$ ``much'' greater than
$r_{0}$ and ``much'' less than $|a|$.  

If we arbitrarily establish a limiting relative error of $10\%$ in the 
comparison between the finite scattering length potential, as plotted
in Fig.~\ref{fig.b}, and the Efimov potential, then we see that
for positive scattering lengths in the range $20r_{0} \lesssim R
\lesssim {|a|}/{13}$ the effective potentials can be considered an
Efimov potential. For $a\le 30000$~a.u., however, the effective
potential does not come within $10\%$ of the Efimov potential in the
range $r_{0}\le R \le a$. For negative scattering lengths, the
effective potentials can be considered (with an error of $10\%$) to be
an Efimov potential in the range $33r_{0}\lesssim R \lesssim
|a|/13$. Thus, both positive and negative $a$ results are reasonably
consistent with the assumption that the Efimov result holds for
$r_{0}\ll R \ll |a|$, but are not consistent with 
$r_{0}\lesssim R \lesssim |a|$. Since the standard estimate for the
number of Efimov states, Eq.~(\ref{NumEfimovStates}), assumes that
the Efimov potential holds in the full range of $R$ from $r_{0}$ to
$|a|$, these more restrictive ranges reduce the estimated number of
Efimov states by about $2$. 

For the $^{4}$He three-body problem, for instance, 
where $a\simeq170$~a.u., the effective potential does not approach
within $10\%$ of the Efimov potential in the range $r_{0}\le R \le
a$. The above-mentioned restrictions lead to the conclusion that no
Efimov state should occur in this case. Nevertheless, accurate
numerical calculations predict a single Efimov state
\cite{HeliumTrimer}, which agrees with the result obtained using
Eq.~(\ref{NumEfimovStates}) where no such restrictions were
imposed. Therefore, despite the more restrictive assumptions discussed
here, Efimov physics can still be observed --- even when the effective
potential does not approach the Efimov potential. The potentials are
still modified in the range $r_{0} \lesssim R \lesssim |a|$, just not
as dramatically. This modification can be seen in Fig.~\ref{fig.b} for
$a=500$, where the curve shows a pronounced maximum near
$R=100$~a.u. before assuming its asymptotic behavior. This sort of
deviation seems to indicate the influence of Efimov physics --- the
smaller this feature, the weaker the influence. 

\begin{figure}[htbp]
\includegraphics[width=2.2in,angle=270]{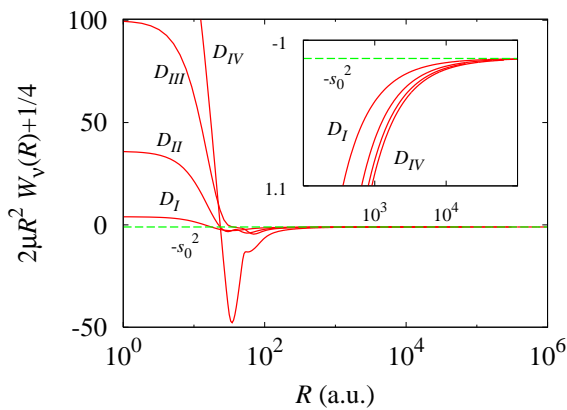} 
\caption{The attractive dipole effective potential for different number
  of two-body bound states. $D_{I}$, $D_{II}$, $D_{III}$ (as given in
  Fig.~\ref{fig.a}) and $D_{IV}$ (not shown in Fig.~\ref{fig.a})
  indicates the effective potentials with no $s$-wave bound states,
  and with one, two and three $s$-wave bound states,
  respectively. \label{fig.b.BS}}
\end{figure}

In Fig.~\ref{fig.b.BS} we show the effective potential for the Efimov
channel with $|a|=\infty$ for systems with different numbers of
$s$-wave bound states. The labels $D_{I}$, $D_{II}$, $D_{III}$ and
$D_{IV}$ denote the two-body potential strengths that produce
$|a|=\infty$ with no $s$-wave bound states, and with one, two and
three $s$-wave bound states, respectively (see Fig.~\ref{fig.a}). For
systems with multiple $s$-wave bound states there are also
rotationally excited two-body bound states that are included in 
the numerical calculations. The inset in Fig.~\ref{fig.b.BS} shows
that the lower limit on the range of validity of the Efimov potential
(\ref{longrange}) changes with the number of two-body bound
states. Therefore, the restriction on the lower limit grows even
stronger when the number of bound states increases. For instance, with 
two $s$-wave bound states ($D_{III}$ in Fig.~\ref{fig.b.BS}), the lower
limit must be set at $48r_{0}$ to meet our $10\%$ relative error
criterion. Figure~\ref{fig.b.BS} also appear to converge to a limiting
curve as the number of two-body bound states increases. 

\begin{figure}[htbp]
\includegraphics[width=2.2in,angle=270]{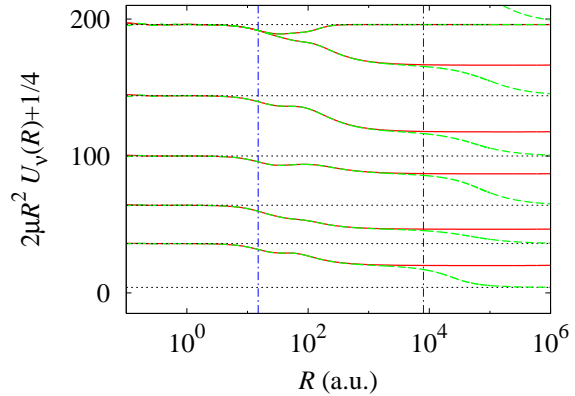} 
\caption{Three-body effective potentials associated with the continuum 
  channels for $|a|=\infty$ (solid lines) and for
  $a=8000$~a.u. (dashed lines). The dotted lines correspond to
  the eigenvalues of the kinetic energy; and the vertical dot-dashed
  lines to the boundaries $r_{0}$ and $a$. \label{fig.b.CC}}
\end{figure}

It is interesting to notice that for $|a|\rightarrow\infty$
the continuum channels also reflect Efimov physics. While they 
still represent three free particles, they also include some
sort of correlation in addition to the kinetic energy. Figure
\ref{fig.b.CC} shows the effective potentials for the continuum
channels with $|a|=\infty$ (solid lines) and
$a=8000$~a.u. (dashed lines). As $R\rightarrow\infty$, the
curves for finite $a$ approach the eigenvalues of the kinetic energy
$\lambda(\lambda+4)+4$ (the dotted lines in
Fig.~\ref{fig.b.CC}). The effective potentials for $|a|=\infty$,
however, converge to a different limit when $R\rightarrow\infty$,
and can be determined analitically \cite{Efimov}. From
Fig.~\ref{fig.b.CC}, we can see that some of the curves for
$|a|=\infty$ still converge to the kinetic energy eigenvalues. 
{Generally, these curves are degenerate in $\lambda$
with $l_{0}\ne 0$ --- only states that have primarily $s$-wave
two-body character are impacted by Efimov physics}. As was the case 
for the attractive dipole potential, the effective
potentials for the continuum channels with finite $a$ behave like the Efimov
potentials [Eq.~(\ref{longrangerep})] in the range $r_{0}\ll R\ll |a|$.
In fact, these modifications to the effective potentials have been
seen for other three-body systems (also for higher partial waves) and
have been shown to play a fundamental role in the scattering length
scaling laws of ultracold three-body collisions \cite{ScaLenLaws}. 

\begin{figure*}[hbtp]
\includegraphics[width=4.5in,angle=270]{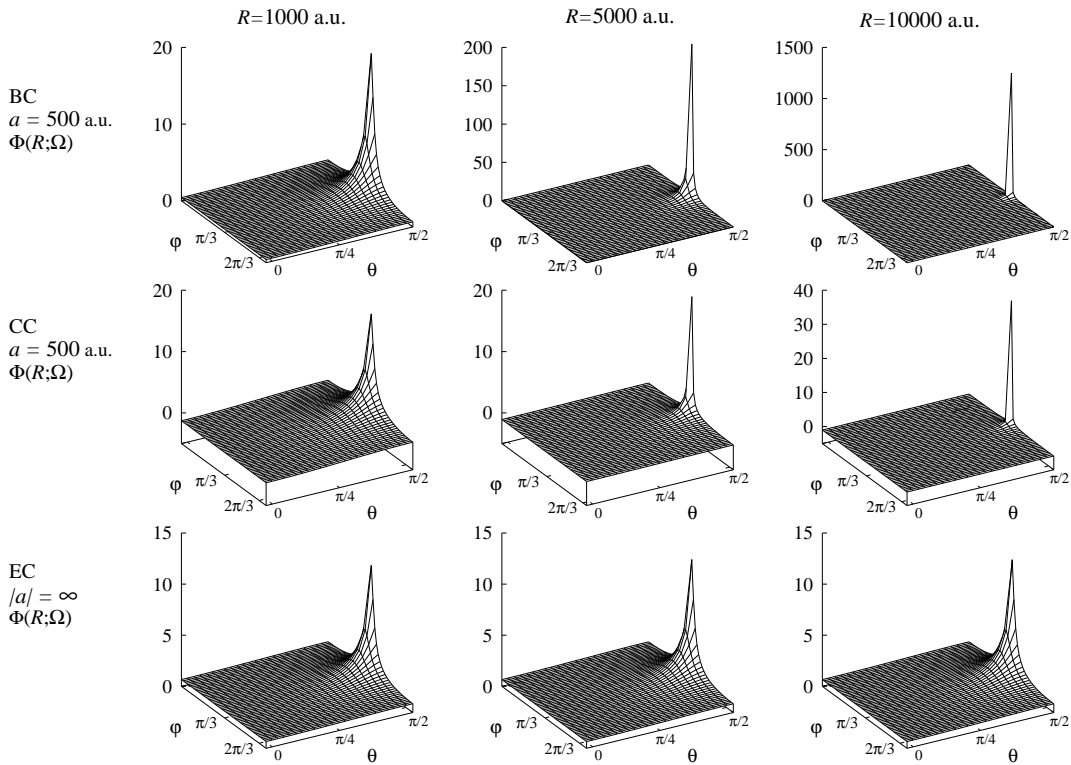} 
\caption{$J^{\pi}=0^{+}$ channel functions $\Phi(R;\Omega)$ for fixed
  values of the hyperradius $R$. The first and second lines show the
  bound channel (BC) and the lowest continuum channel (CC) for
  $a=500$~a.u.. The third line shows the Efimov 
  channel (EC), demonstrating that it is insensitive to changes in
  $R$. \label{fig.c}}   
\end{figure*}

\subsection{Channel Functions}

In Fig. \ref{fig.c} we show the channel functions for the three
classes of states: bound (BC), continuum (CC), and Efimov (EC). 
For the $J^{\pi}=0^{+}$ symmetry, the channel functions do not
depend on the Euler angles; they can thus be visualized for fixed
values of $R$ as a function of the hyperangles $\varphi$ and $\theta$
only. We show only one third of the range of $\varphi$ in
Fig.~\ref{fig.c} since the rest of the wave function can be obtained
by symmetry: 
$\Phi(\varphi+4\pi/3)=\Phi(\varphi+2\pi/3)=\Phi(\varphi)$. Further, we
show the region that includes $r_{13}=0$ --- this point lies at
$\varphi=\pi/3$ and $\theta=\pi/2$. The hyperangular distribution 
tells us about the geometry of the system. At $\theta=0$, for
instance, the atoms form an equilateral triangle and at $\theta=\pi/2$ 
they lie along a line.

As $R$ increases, a bound channel function collapses like $R^{-1}$ to
the region around $r_{ij}=0$, as shown in the first row of
Fig. \ref{fig.c} for 
$a=500$~a.u.. The channel function occupies a smaller and smaller
region of the ($\varphi,\theta$) plane as $R$ increases, so must grow
in amplitude to preserve normalization. The second row of
Fig. \ref{fig.c} represents the lowest continuum channel function
for $a=500$~a.u., which is the second
channel for the system. Therefore, the channel function must approach
the $\lambda=0$ hyperspherical harmonic (a constant function of the
hyperangles) as $R\rightarrow\infty$, and it must have one
node (careful inspection of the figure near $\theta=0$ shows that
$\Phi<0$). In   
Fig. \ref{fig.c} the harmonic character is easily seen since the
channel function is spread nearly uniformly over the entire
hyperangular plane. It is also apparent that there is a substantial
amplitude at the two-body coalescence point and that this amplitude
shrinks with growing hyperradius just as for the bound channel. 

The last row of Fig. \ref{fig.c} shows the channel function for 
$|a|=\infty$. These channel functions share characteristics of
both the bound and continuum channels. That is, they have large
amplitude at the two-body coalescence points, but they are also spread 
over the entire plane. These channel functions, however, have one
important characteristic that the others do not have: they are nearly
scale invariant. That is, they look essentially the same at every
value of $R$. This scale invariance is, in fact, one of the defining
properties of the Efimov effect. These channel functions also show why
the Efimov effect is so tenuous --- the three bodies must maintain a
delicate balance between the two-body and three-body parts of
configuration space over ever-growing volumes of space. {If there are
two-body bound states present, this Efimov channel function will
aquire nodes closely resembling those of the continuum channel.}  

\begin{figure}[htbp]
\hspace{-1.cm}\includegraphics[width=2.6in,angle=270]{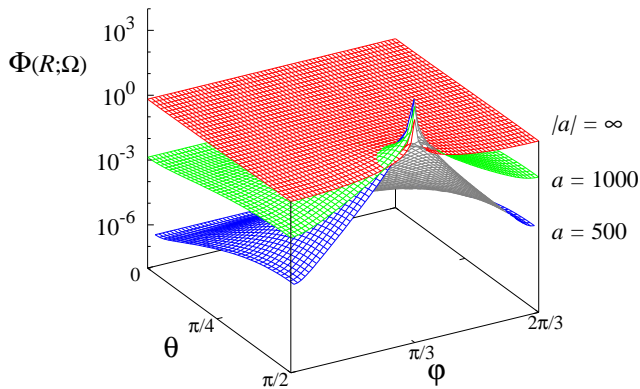} 
\caption{Channel function for the weakly bound channel at
  $R=10000$~a.u. for $a=500$~a.u. and $a=1000$~a.u.. The $|a|=\infty$
  channel function is also shown for comparision. Note that the
  vertical scale is logarithmic. \label{fig.c.together}} 
\end{figure}

In Fig.~\ref{fig.c.together} we compare the channel functions for the
weakly bound channel for $a=500$~a.u., $a=1000$~a.u., and
$|a|=\infty$ at $R=10000$~a.u.. This figure shows the channel function
on a logarithmic scale (in the $z$-direction) to emphasize the
delocalization in the Efimov case ($|a|=\infty$). At this value
of $R$, the delocalization grows with $a$, while all cases have
large amplitude at the two-body coalescence points.  

\subsection{Nonadiabatic couplings}

The asymptotic behavior of the nonadiabatic coupling for finite $a$
has been deduced analytically for all combinations of bound and
continuum channels \cite{Nielsen}. The coupling to the Efimov channel,
however, was only briefly discussed \cite{Fedorov}, and that was only
for the diagonal coupling element $Q_{\nu\nu}$. We have found that the
nonadiabatic coupling is also modified due to Efimov physics in the
range $r_{0} \ll R \ll |a|$. In Table~\ref{tab.a.new} we show the
leading term in the asymptotic expansions of $P_{\nu\nu'}$ and
$Q_{\nu\nu'}$ couplings deduced in Ref.~\cite{Nielsen} which are
expected to be valid for $R\gg |a|$. Here, we will restrict our
analysis to couplings between $s$-wave channels and the Efimov
channel. The numerically calculated couplings were obtained for
values of the interaction strength near $D=D_{II}$ (Fig.~\ref{fig.a}),
where the system has one bound state for $a<0$ and two bound states for  
$a>0$. In this case, therefore, we investigate the behavior of the
nonadiabatic coupling in a situation similar to crossing an $s$-wave
Feshbach resonance \cite{Inouye,Courteille,Roberts,Kevrekidis,Weber},
which involves the formation of a weakly bound molecule for $a>0$.
\begin{table}[htpb]
\begin{tabular}{c|cc}\hline\hline
$\nu$$-$$\nu'$ & $P_{\nu\nu'}(R)$ & $Q_{\nu\nu'}(R)$ \\ \hline
BC($l=0$)$-$BC($l=0$) & $R^{-1}$ & $R^{-2}$  \\
BC($l=0$)$-$CC($l_0=0$) & $R^{-5/2}$ & $R^{-7/2}$   \\
CC($l_0=0$)$-$CC($l_0=0$) & $R^{-2}$ & $R^{-4}$ \\ 
\hline \hline
\end{tabular} 
\caption{The leading term for the asymptotic expansion of the
  nonadiabatic coupling involving bound channels (BC) and continuum
  channels (CC). The diagonal coupling $Q_{\nu\nu}$
  for BC is proportional to $R^{-2}$, while for CC it is proportional
  to $R^{-4}$ (see Ref.~\cite{Nielsen}).} \label{tab.a.new}
\end{table}

\begin{figure}[htbp]
\begin{center}
\includegraphics[width=3.9in,angle=270]{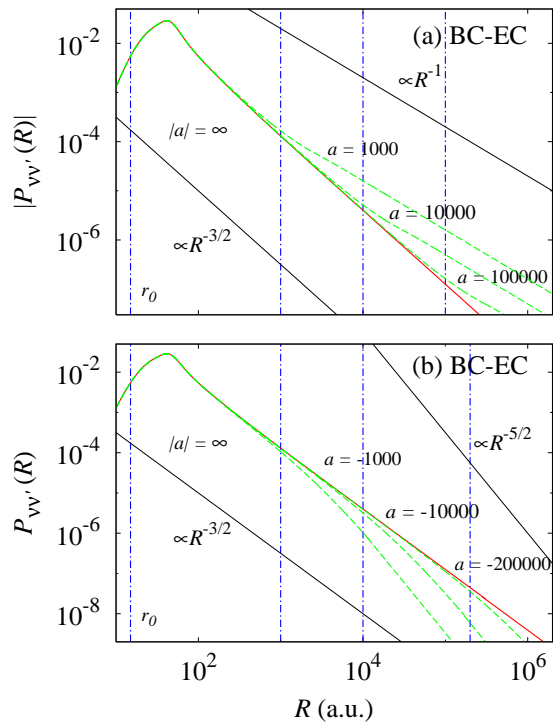} 
\end{center}
\caption{The nonadiabatic couplings between a bound channel and the
  Efimov channel for finite (dashed lines) and infinite (solid line)
  scattering lengths.   For $a>0$ (a) the Efimov channel is associated
  with the weakly bound channel; and for $a<0$ (b) with the lowest
  continuum channel. \label{fig.d}}
\end{figure}

\begin{figure}[htbp]
\begin{center}
\includegraphics[width=3.9in,angle=270]{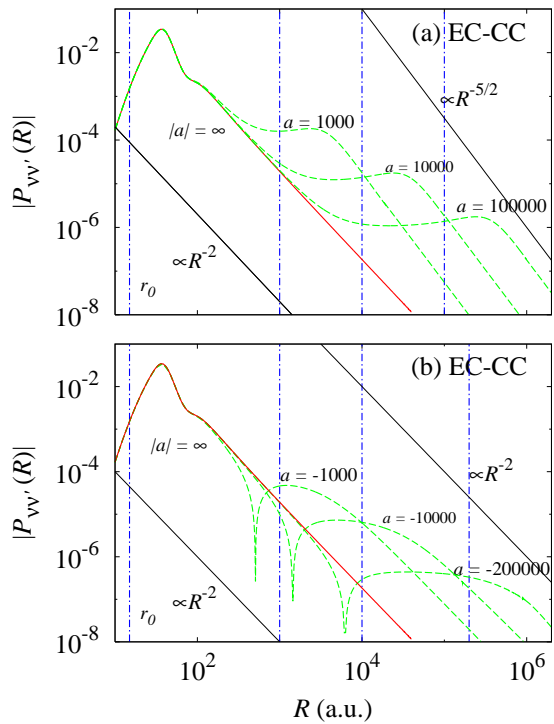} 
\end{center}
\caption{The nonadiabatic couplings between the Efimov channel and a
   continuum channel for finite (dashed lines) and infinite (solid
   line) scattering lengths. For $a>0$ (a) the Efimov channel is
   associated with the weakly bound channel; and for $a<0$ (b) with
   the lowest continuum channel. \label{fig.e}} 
\end{figure}

\begin{figure}[htbp]
\begin{center}
\includegraphics[width=3.9in,angle=270]{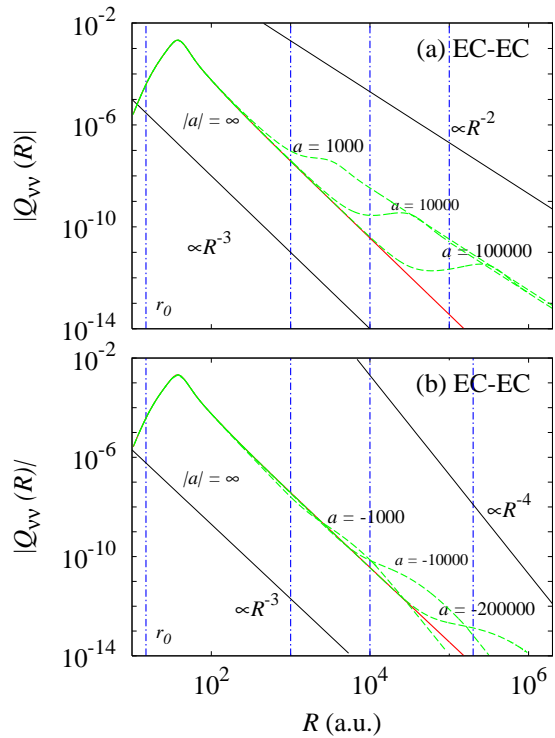} 
\end{center}
\caption{The diagonal nonadiabatic coupling for the Efimov channel for
  finite (dashed lines) and infinite (solid line) scattering
  lengths. For $a>0$ (a) the Efimov channel is associated with the
  weakly bound channel; and for $a<0$ (b) with the lowest continuum
  channel. \label{fig.f}}
\end{figure}

We plot in Figs. \ref{fig.d}, \ref{fig.e}, and \ref{fig.f} the
numerically calculated nonadiabatic couplings on a log-log scale in
order to highlight the expected power law behavior. Figure \ref{fig.d}
shows the couplings between an $s$-wave bound channel and the Efimov
channel for finite $a$ (dashed lines) and $|a|=\infty$ (solid line).
For $a>0$, and $R\gg a$, the couplings behave as couplings between
bound channels, given in Table~\ref{tab.a.new}.
For $a<0$, they behave as couplings between bound and continuum
channels. As $|a|\rightarrow\infty$, the couplings in both
Figs.~\ref{fig.d}(a) and (b) approach a limiting value given by the
$|a|=\infty$ result. In the range $r_{0}\ll R \ll |a|$, indicated in
Fig.~\ref{fig.d} by the vertical dot-dashed lines, the couplings are
modified due to Efimov physics. 
The $|a|=\infty$ coupling at the other poles of Fig.~\ref{fig.a} will
have the same behavior for $R\gg r_0$, but the short-range behavior
will differ.

Figure~\ref{fig.e} shows the coupling between the Efimov channel
and the lowest continuum channel for $a>0$ and the next continuum
channel for $a<0$. 
The $a>0$ curves in Fig.~\ref{fig.e}(a) show the characteristic peak
around $R\simeq 3a$ predicted in \cite{Macek-02} responsible for
three-body recombination at ultracold temperatures. For $a<0$, there
is one node at the dip in Fig.~\ref{fig.e}(b) (since the absolute
value was taken for this plot). As $|a|$ increases, the node moves to
infinity, and the coupling for the Efimov case emerges.  

The asymptotic behavior of the diagonal correction in
Fig.~\ref{fig.f} also confirms the results in Ref.~\cite{Nielsen} for
finite $a$. In Fig.~\ref{fig.f} the diagonal coupling for
$|a|=\infty$ is proportional to $R^{-3}$, which lies between the
$R^{-2}$ expected for $a>0$ and $R^{-4}$ expected for $a<0$ (see
Table~\ref{tab.a.new}). 

Our calculations confirm all of the finite $|a|$ asymptotic
analysis of Ref.~\cite{Nielsen}. Table \ref{tab.a} summarizes the
cases for coupling with the Efimov channel when $|a|=\infty$. It shows
the leading term in the expansion of the nonadiabatic couplings for
three identical bosons and are expected to be valid in the range
$r_{0} \ll R \ll |a|$, as with the other results related to Efimov
physics. Since the continuum channels are also modified due to the
Efimov physics (see Fig.~\ref{fig.b.CC}), the couplings between bound
channels and the continuum channels are expected to change. In fact,
only the coupling between $s$-wave bound channels and continuum
channels with $l_{0}=0$ are different from the finite $a$ case. We 
have included this coupling in the last line of Table \ref{tab.a} for
completeness. 
\begin{table}[htpb]
\begin{tabular}{c|cc}\hline\hline
$\nu$$-$$\nu'$   & $P_{\nu\nu'}(R)$ & $Q_{\nu\nu'}(R)$ \\\hline  
EC$-$BC($l = 0$) & $R^{-3/2}$ & $R^{-5/2}$     \\
EC$-$CC($l_0=0$) & $R^{-2}$   & $R^{-3}$  \\
BC($l=0$)$-$CC($l_0 = 0$) & $R^{-3/2}$   & $R^{-5/2}$  \\ 
\hline \hline
\end{tabular}
\caption{The leading term for the asymptotic expansion of the
  nonadiabatic couplings for $|a| = \infty$. The diagonal coupling
  $Q_{\nu\nu}$ for the Efimov channel is proportional to $R^{-3}$. 
  \label{tab.a}}    
\end{table}

Comparing Tables~\ref{tab.a.new} and \ref{tab.a}, we
see that the Efimov channel shares characteritics of bound
and continuum channels. The coupling between the Efimov channel and
bound channels ($R^{-3/2}$) falls faster than the coupling between 
bound channels ($R^{-1}$), but slower than the coupling between 
bound and continuum channels ($R^{-5/2}$). The coupling between the
Efimov channel and continuum channels ($R^{-2}$) falls faster
than the coupling between bound and continuum channels
($R^{-5/2}$), but, is the same as the coupling between continuum
channels ($R^{-2}$). 

From Figs.~\ref{fig.d}, \ref{fig.e} and \ref{fig.f} it is clear
that the couplings for finite $a$ also exhibits universal
aspects. This universal behavior would better seen by plotting
$|a|P_{\nu\nu'}(R/|a|)$ and $|a|^{2}Q_{\nu\nu'}(R/|a|)$. In
such a plot the couplings for different values of $a$ look all 
the same for $R/|a|\gg 1$.

\section{Summary \label{secIV}}

We have investigated several fundamental aspects of Efimov physics in
the hyperspherical adiabatic representention of systems of three
identical bosons. We have calculated effective 
hyperspherical potentials for very large to infinite values of the
two-body $s$-wave scattering length. These calculations were 
obtained in a regime not accessible in previous calculations, allowing
us to test assumptions related to the Efimov effect, namely, the
emergence of the long-range three-body effective interactions in the
range where $r_{0}\ll R\ll |a|$. 
Our results show that despite the more restrictive limits discussed
here for the effective potentials, the assumption that the long-range
attractive potential holds in the range $r_{0}\lesssim R\lesssim |a|$
gives a good qualitative description of the Efimov physics. In fact,
for $a\gtrsim r_{0}$, the unrestricted result for the number of Efimov
states seems to give a better description, even though the effective
potential does not resemble the Efimov potential. 
Additionally, we have shown that
as we increase the number of two-body bound states the effective
potential seems to converge to a limiting curve.

We have also obtained the behavior of many other quantities affected
by Efimov physics: the continuum effective potentials, all of the
nonadiabatic couplings with the Efimov channel, and the channel
functions themselves. As expected, the channe functions are
essentially unchanged with $R$. Moreover, the nonadiabatic couplings
reveal that the Efimov channel behaves in some ways like a bound state
and in others like a continuum state. We have quantified these
different regimes and verified that the effects are limited to
$r_{0}\ll R \ll |a|$ for all quantities.

\acknowledgments{This work was supported by the National
 Science Foundation and by the Research Corporation.}

\end{document}